\documentclass[debug,overfull,preprint,figures,cite,bm]{epl}
\usepackage{epsf}
\usepackage{psfig}
\usepackage{amssymb}
\usepackage{amsfonts}
\usepackage{amsmath
}

\title{Detachment of molecular motors under tangential
loading}
\shorttitle{Detachment of ...}

\author{A.
Parmeggiani\inst{1,2,3}
\and F. J\"ulicher\inst{1}
\and L.
Peliti\inst{4} \and J. Prost\inst{1} }

\institute{
            \inst{1}
Institut Curie, Physico-Chimie Curie, UMR 168
CNRS/IC,
26 rue d'Ulm,
F--75248 Paris Cedex 05, France \\
            \inst{2} Institut f\"ur
Theoretische Physik, T34, Physik
Department, TU-M\"unchen,
James-Franck-Stra\ss e, D-85747 Garching, Germany \\
 \inst{3}
Lyman Laboratory of Physics,
Harvard University,
Cambridge MA 02138,
USA
 \\
\inst{4} Dipartimento
di Scienze Fisiche and Unit\`a INFM,
Universit\`a "Federico II", Complesso
Monte S. Angelo, I--80126
Napoli, Italy }

\date{January 25,
2000}

\begin{document}
\maketitle

\begin{abstract}
We introduce a
general framework to study the processivity of
molecular motors moving
along a polar filament and discuss the average
time spent attached to the
filament as a function of a tangentially
applied load.  Our study of
specific models suggests that the
attachment time of a motor decreases with
increasing ATP concentration
and that double-headed motors such as kinesins
lose their processivity
under forcing conditions while processive
single-headed motors are
less sensitive to tangential
forcing.
\end{abstract}

\section{Introduction}

Experiments on motor
proteins crucially depend on an important
property of such systems, called
processivity.  Motor proteins such as
kinesins, which can hydrolyse a large
number of ATP molecules before
detaching from the microtubule, are called
processive, whereas those
like myosin~II, which typically interact just
once with actin
filaments and detach, are called non-processive
\cite{howa97}.
Physiologically, processivity is important since it permits
transport
involving a small number of motors.  Non-processive motors,
like
myosin~II in muscles, are found to work in large assemblies.
After
having collected important experimental information on
elementary
steps, the velocity as a function of load and the stall force
of
molecular motors such as kinesin \cite{svob93,meyh95,viss99}, there
are
now efforts to quantitatively study the processivity.  As a
result, there
is now ample experimental evidence that applying a force
tangential to the
cytoskeletal filaments and opposing the natural
motion of molecular motors
considerably decreases the processivity
\cite{viss99,copp97,schn00}.

A
priori, one can define processivity in three different ways.  First
it can
be chosen as the average number of chemical cycles before
detachment from
the filament; second it can be defined as the
attachment lifetime of the
motor to the filament; third, it can be, as
in Ref.  \cite{schn00}, chosen
to be the mean length spanned by the
motor on the filament in a single run.
The first definition is
intrinsic to the process but extremely difficult to
measure.  The
other two are directly accessible to experiment.  For the
sake of
conciseness we focus our discussion on the attachment lifetime
and
only give an example for the run length.

The existence of a disruptive
force normal to the cytoskeletal
filament could potentially explain the
decrease in processivity, since
it is clear that tangential forces applied
on a bead or on a rod
attached to the motor produce a normal component on
it.  The
detachment produced by normal forces is similar to the detachment
of
adhesion proteins under load, investigated
in
Refs.~\cite{Evan97,merk99}. However, in the case of
processive molecular
motors there is an additional physical mechanism
leading to detachment due
to applied tangential forces. This effect is
the subject of the present
letter.

Our calculations show that applying a force strictly tangential to the
filament in general reduces
the average attachment time of molecular
motors. This effect is
particularly pronounced for a model motivated
by double-headed processive
motors, which provides a natural
explanation for the loss of kinesin
processivity observed under
loading conditions \cite{viss99,copp97,schn00}.
It is interesting to
remark
that for a model adapted to single-headed
processive motors we find a
less marked force-dependence of the average
attachment time.

\section{Model}
These results are obtained by taking into
account the possibility of
detachment in $N$-state models of molecular
motors
\cite{ajda92,astu94,pros94,oste94,chau94,astu97,juli97,parm99}.
In
these models, the motor is represented by a point particle which can
be
in one of $N$ internal states (representing the steps of the
chemical
cycle) and is placed at point $x$ in the one-dimensional
space representing
the filament. For each state $i$, the particle is
subject to a periodic
potential $W_i(x)$ of period $l $. The
transition rates $\omega_{ji}(x)$
depend on ATP concentration as
discussed in Ref. \cite{parm99}.  Its
hydrolysis drives the motion
along the filament.

Detachment can be
represented in this picture by allowing the particle
to disappear with a
local detachment rate $\alpha_i(x)$ which depends
both on its state and on
its position. It is reasonable to expect that
$\alpha_i(x) \sim
\exp(W_i(x)/k_{\rm B}T)$, since the ``attempt
rates'' for detachment should
not depend very strongly on position.
The exponential dependence is so
strong that, to a first
approximation, one may consider high detachment
rates to be
concentrated near the maxima of the potential; elsewhere we
assume a
constant, much smaller, detachment background.

We assume that the
particle attaches to the filament at $t=0$ with a
probability distribution
$P^{(0)}_i(x)$.
The behavior of the system can be described by
the
probability density $P_i(x,t)$ to find the particle at position $x$
and
state $i$ at time $t$. This quantity satisfies, for $t>0$, the
evolution
equation
\begin{equation}
\partial_t P_i+\partial_x J_i=\sum_j
\left(\omega_{ij}P_j
-\omega_{ji}P_i\right)-\alpha_i
P_i,
\end{equation}
where the currents $J_i$ are defined
by
\begin{equation}
J_i=-\mu_i \left[k_{\rm B}T\partial_x
P_i+\left(\partial_x W_i-
f_{\rm ext}\right)P_i\right],
\end{equation}
in
which $\mu_i$ is the particle mobility and $f_{\rm ext}$ the
externally
applied tangential force.

Since the probability that the particle detaches
anywhere on the
filament between time $t$ and $t+d t$ is given
by
$\int_{-\infty}^{+\infty} dx\, \sum_i \alpha_i(x)P_i(x,t)$, we
can
evaluate the average attachment time $\tau=\int_0^\infty d
t\,
t\int_{-\infty}^{+\infty}d x\, \sum_i \alpha_i(x)P_i(x,t)$.  It
is
straightforward to show that this definition is equivalent
to
\begin{equation}
\tau=\int_0^\infty d t\int_{-\infty}^{+\infty}d
x\,\sum_i P_i(x,t).
\end{equation}

Similarly we can define the mean run
length as
\begin{equation}
\lambda=\int_0^\infty d
t\,
\int_{-\infty}^{+\infty}d x\, x \sum_i \alpha_i(x)P_i(x,t)
\quad.
\end{equation}
Introducing the Laplace transform $\tilde
P_i(x,s)=\int_0^\infty
dt\,e^{-st}P(x,t)$, we
obtain
\begin{equation}
\tau=\int_{-\infty}^{+\infty}d x
\sum_i\tilde
P_i(x,s{=}0) \quad ,
\end{equation}
where $\tilde P_i(x,s)$ is, for each
$s$, a solution of the ordinary
differential equation
\begin{equation}
s
\tilde P_i + \frac{d \tilde J_i}{d x} + \sum_j (\omega_{ji}
\tilde
P_i
-\omega_{ij} \tilde P_j)+\alpha_i \tilde P_i
=
P^{(0)}_i(x),\label{fund:eq}
\end{equation}
which we only need to solve
for $s=0$.

\section{Method}

Since eq.~(\ref{fund:eq}) is linear, we can
solve
it for $P^{(0)}_i(x)=\delta_{ii_0}\delta(x-x_0)$
without loss of
generality.
This implies that a motor is initially
placed at position
$x_{0}$ and in state $i_{0}$. In our numerical
analysis, we choose
$x_{0}=0$ and $i_{0}=1$.
Then, for
$x>x_0$ and $x<x_0$, the solution
can be found
by a transfer matrix technique. The initial solution
imposes
the matching conditions
\begin{eqnarray}
P_i(x_0+\epsilon,s)&=&P_i(x_0-\epsilon,s)
\nonumber\\
J_i(x_0+\epsilon,s)&=&J_i(x_0-\epsilon,s)+\delta_{ii_0}\label{incond:eq}.
\end{eqnarray}
Assuming that the vector
$\vec
Y=(y_1,y'_1,\ldots,y_N,y'_N)$ contains
the values of the solution and its
first derivatives at point $x$,
one can evaluate the
solution at point
$x+nl $
as ${\cal M}^n\vec Y$, where the transfer matrix
$\mathcal{M}$ is
evaluated by explicitly solving eq.~(\ref{fund:eq})
within a period.
Notice
however that, since the boundary conditions
imply that $P(x,t)$ vanishes as
$x \to \pm\infty$ at all times, this
must be also true for $\tilde P(x,s)$
for all $s$.  Therefore,
when $x>x_0$ ($x<x_0$),  $Y$ must
belong to the
$N$-dimensional
space spanned by the eigenvectors of $\mathcal{M}$
(${\cal M}^{-1}$) whose absolute value is smaller
than 1. The $2N$ matching
conditions (\ref{incond:eq})
are thus in general sufficient to fix the $2N$
coefficients
which identify the solution.

The evaluation of the transfer
matrix is difficult in general, but can
be made tractable by assuming that
the $\omega_{ij}(x)$ and the
$\alpha_i(x)$ are piecewise constant, and that
the $W_i(x)$ are
piecewise linear. Then the coefficients of
eq.~(\ref{fund:eq}) are
piecewise constant, and the solution can be
expressed by exponentials,
which must be smoothly connected at the joining
points, ensuring
the continuity of the probability density $P$ and of the
current $J$.
Therefore, at the price of a special treatment for the
period
containing $x_0$, the solution of eq.~(\ref{fund:eq}) reduces to
a
set of transcendental equations whose solution can be
numerically
evaluated~\cite{thesis,long}.
In particular, the infinite-force
limit~\cite{thesis}
and the simplest case with only one state
can be fully
solved analytically~\cite{long}.

\section{Examples}

\begin{figure}
\twoimages[scale=0.4]{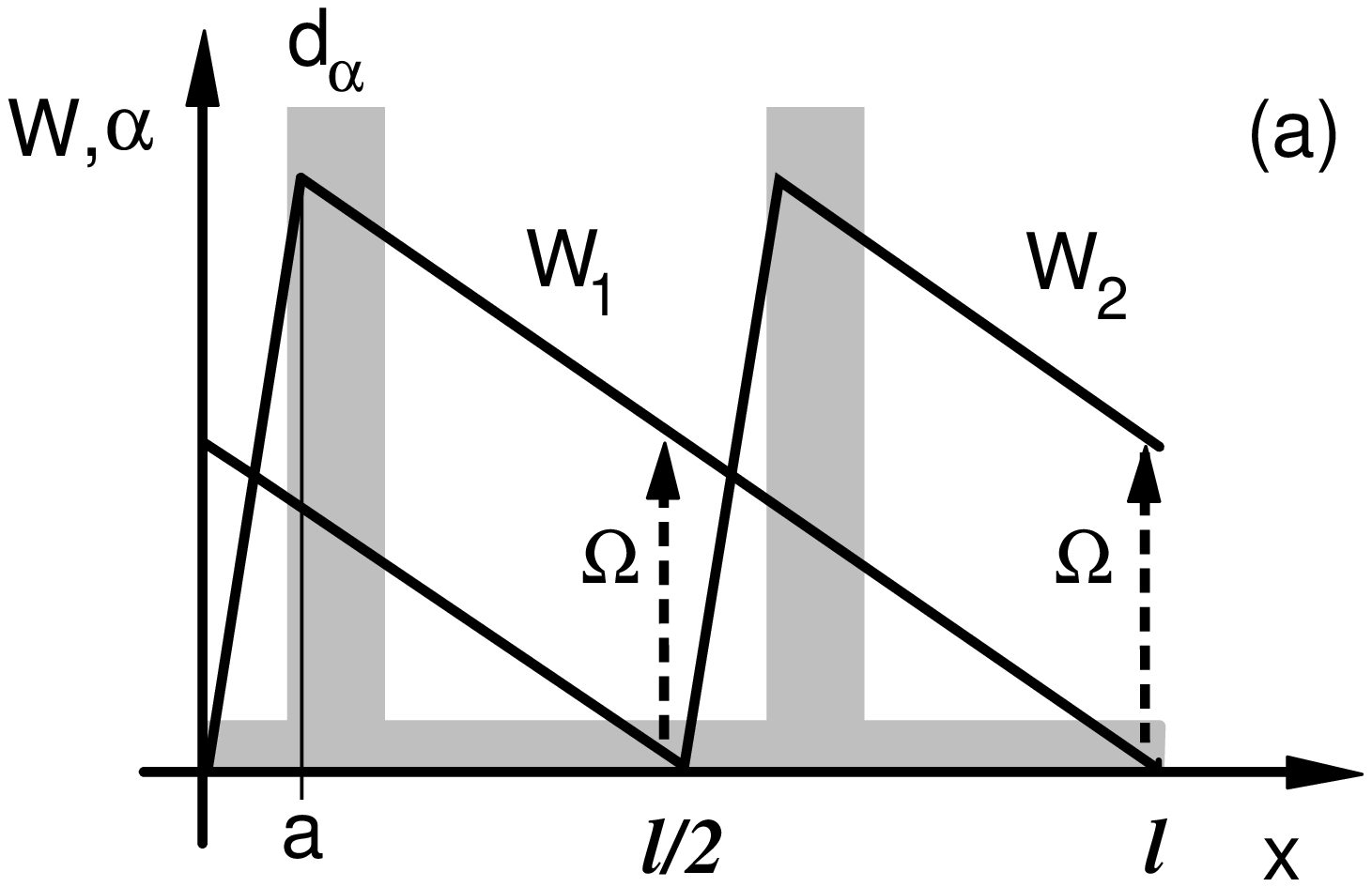}{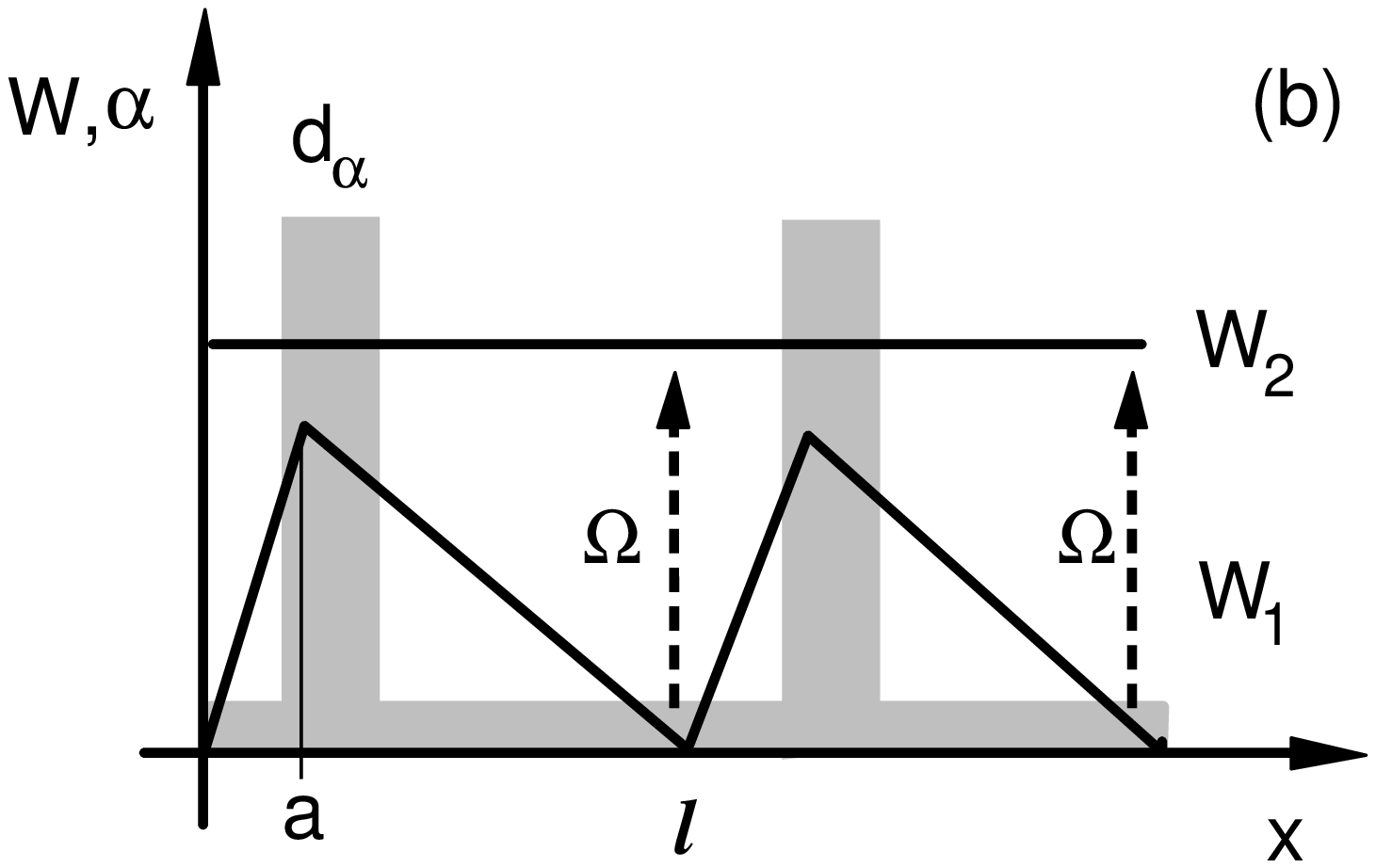}
\caption{Schematic representation of two specific models discussed
in the text.  (a)
Identical shifted states (ISS): Transitions are
localized in a region of
size $d_{\omega}$. The detachment rates
$\alpha$
are piecewise constant
with the maximal value $\alpha_{\rm max}$
localized
in a region of size
$d_\alpha$. Elsewhere $\alpha=\alpha_{\rm min}$.
(b) Diffusive steps (DS):
Excitations are localized at the
minima of $W_1$, deexcitations are
homogeneous. Detachment rates
$\alpha$ are maximal ($\alpha_{\rm max}$)
near the potential maximum and
$\alpha_{\rm min}$ otherwise. The potential
amplitudes are denoted by
$U$, the asymmetry of the sawtooth potentials is
characterized by the
length $a$, and $l$ denotes the potential period. The
value of $W_2$
can satisfy $W_{1\rm min} < W_2 < W_{1\rm max}$, but the
line has been
displaced upwards for graphical
clarity.}
\label{f:pot}
\end{figure}

We discuss two simple
examples with $N=2$ and asymmetric sawtooth
potentials: (i) {\it Identical
shifted states} (ISS): This system
involves two identical potentials
shifted by half a period
($W_1(x)=W_2(x+l /2)$); (ii) {\it Diffusive steps}
(DS): In this case
we choose a flat excited state ($W_2(x)={\rm const.}$)
and all
transition rates for this state are constant.
This simplified
situation is valid as long as the variations of
$W_{2}$ do not exceed $kT$
\cite{chau94,pros94}.
The potentials and the
transition rates for these
situations ISS and DS are defined in
Fig.\ref{f:pot}.  We denote the
potential amplitude by $U$ and
characterize the potential asymmetry by
$a/l$, where $a$ denotes the
position of the maximum of
$W_1$.

The ISS model is motivated by the
motion of double-headed
kinesins. Each of the two states could correspond
to a situation where
one of the heads is strongly bound to the filament.
Because the two
kinesin heads are identical, the two states are identical
up to a
relative shift. We choose transitions between the states
localized
within a region of size $d_\omega$ near the potential minima. At
the
minimum of $W_1(x)$, excitations occur with a rate
$\omega_{21}=\Omega$
while $\omega_{12}=\omega$. Because of the
symmetry of the states $\omega_{21}=\omega$ and $\omega_{12}=\Omega$
at the minima of
$W_2(x)$. Elsewhere, $\omega_{12}$ and $\omega_{21}$
vanish. The rates of
detachment from the filament $\alpha_i(x)$ are
chosen piecewise constant
with a large value $\alpha_{\rm max}$
located in a region defined by
$W_i(x)>U-k_{\rm B} T$ of size
$d_\alpha$
around the potential maxima, and
with a small value $\alpha_{\rm min}$
elsewhere, see Fig.\ref{f:pot}(a).

Note, that this simple choice captures the essential exponential
dependence
of the detachment rate but is more convenient to use for
practical
calculations.

The use of the DS model is motivated by the
recently observed
processive motion of artificially constructed individual
kinesin heads
\cite{okad99,okad00}. The particle, initially lying
in the
potential minima in state 1, gets excited into state 2, where
it diffuses.
It then falls back into state 1, and advances if
diffusion has brought it
past the nearest potential maximum.  For this
system, we choose excitations
that are localized in the potential
minimum where $\omega_{21}=\Omega$,
$\omega_{21}=0$ elsewhere. The
deexcitation rate $\omega_{12}=\omega$ is
constant along the
period. The detachment rates $\alpha_1(x)$ are
large
($\alpha_1=\alpha_{\rm max}$) in a region of width $d_\alpha$
around
the potential maxima, and have a small value
($\alpha_1=\alpha_{\rm
min}$) elsewhere. We choose $\alpha_2$ to be a
constant, related
to the value of the potential energy in state 2.
The fact
that the observed attachment times of an artificially
constructed
single-headed kinesin is similar to those characteristic
for double-headed
kinesins~\cite{okad99} suggests that in this case
the potential energy of
state 2 is not larger than the potential
maximum of state 1, and thus that
$\alpha_2 < \alpha_{\rm max}$.  If
the potential energy of state 2 is
large, one has $\alpha_2 \ge
\alpha_{\rm max}$, and the corresponding motor
is non processive:
indeed, it is likely to leave the filament as soon as it
gets
excited. This situation could apply, therefore, to
myosin~II.

\section{Results}

\begin{figure}
\onefigure[scale=0.5]{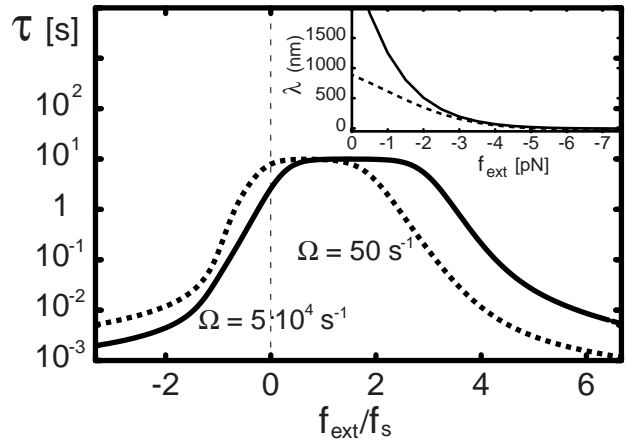}
\caption{Average
attachment time to a filament $\tau$ as a function of
the normalized force $f_{\rm ext}/f_{\rm s}$, where $f_{\rm s}$ is a
typical value of
the stall force. ~Identical
shifted states (ISS) with $U/k_{\rm B}T=20$,
$l=16$~nm, $a/l=0.2$,
$\mu_1=\mu_2=171$~nm~pN$^{-1}$s$^{-1}$,
$\omega=10^{-4}$~s$^{-1}$,
$d_\omega/l =0.02$, $d_\alpha/l =0.05$,
$\alpha_{\rm
min}=0.1$~s$^{-1}$, $\alpha_{\rm max}/\alpha_{\rm min}=4.8\,
10^{8}$
for $\Omega=5.0\;10^4$~s$^{-1}$ and $\Omega=50$~s$^{-1}$.
For
these parameter values, the system has, for large $\Omega$, a
saturated
spontaneous average velocity $v_0=1.1\; \mu$m/s in the
absence of external
forces and the force is normalized with $f_{\rm
s}= 6.0$~pN. Inset: mean
run length as a function of the external
force $f_{\rm ext}$, in the ISS
with identical parameters. Note the
similarity with experimental results of
Ref.
[6].}
\label{f:resA}
\end{figure}

\begin{figure}
\twoimages[scale=0.4]{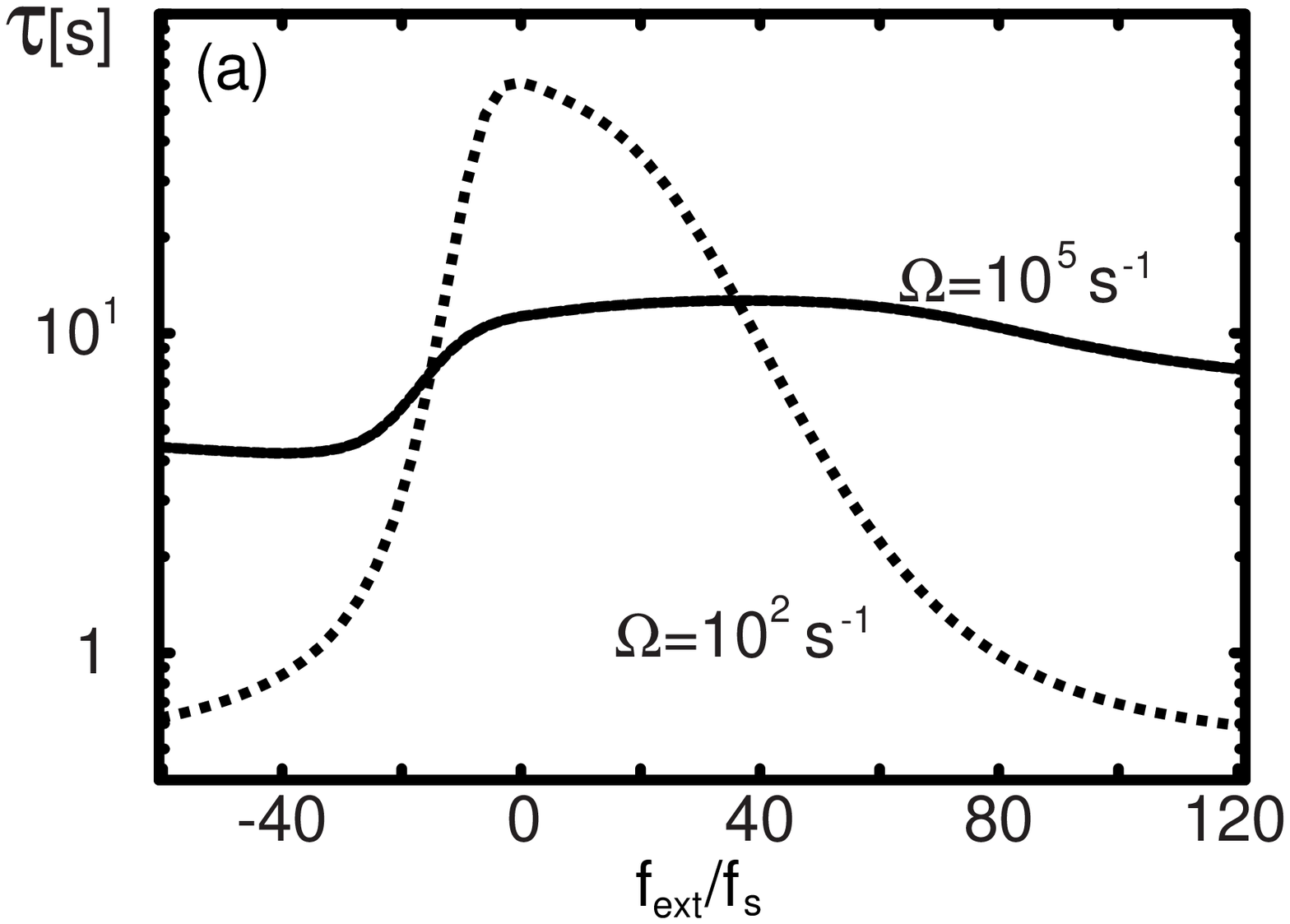}{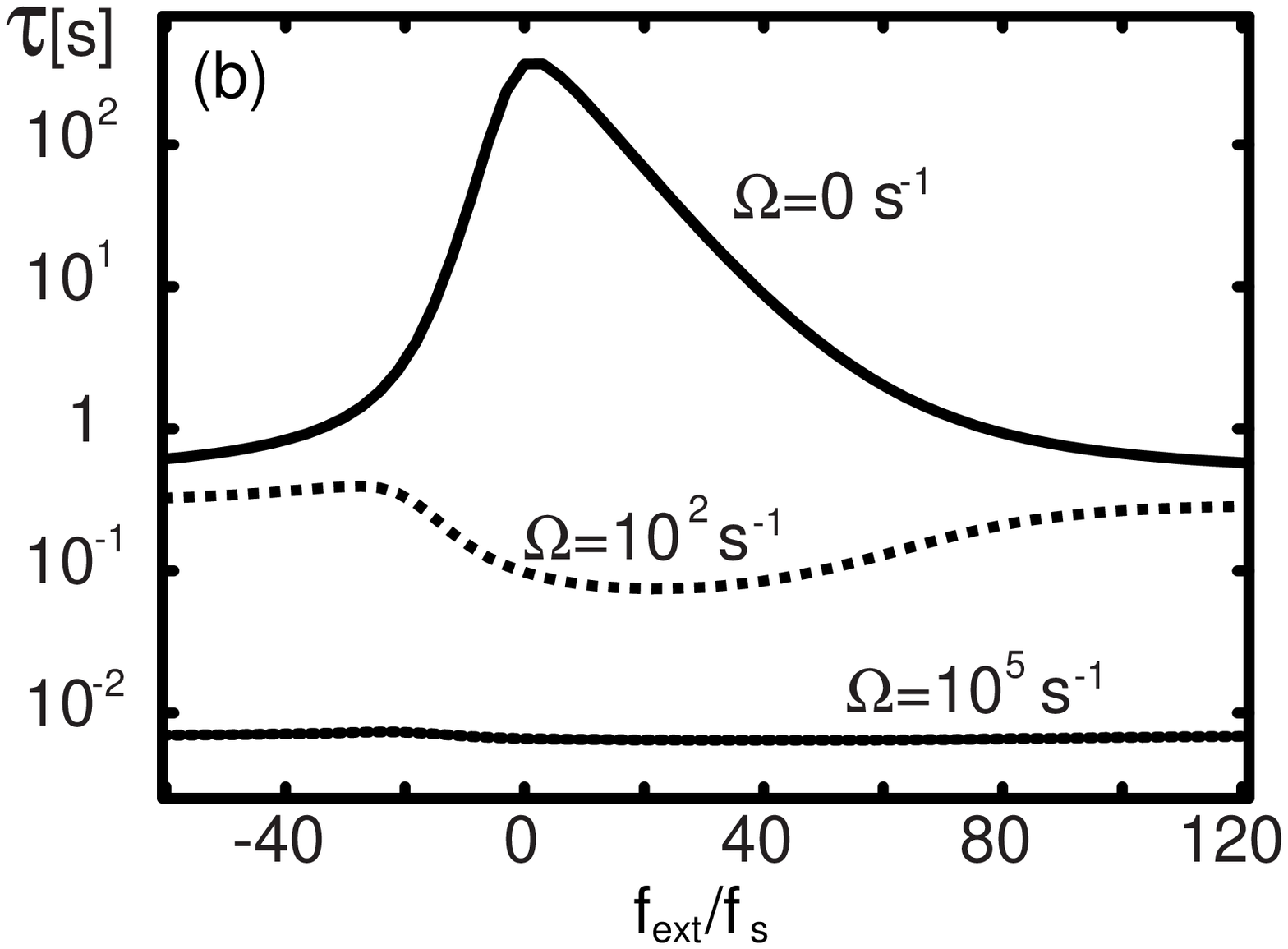}
\caption{(a) Average attachment time to a filament $\tau$ as a
function of
the normalized force $f_{\rm ext}/f_{\rm s}$, where $f_{\rm s}$
is a
typical value of the stall force. Diffusive steps (DS) with
$U/k_{\rm
B}T=10$, $l =8$~nm, $a/l=0.2$,
$\mu_1=\mu_2=5\;10^2$~nm~pN$^{-1}$s$^{-1}$, $\omega=100$~s$^{-1}$,
$d_\omega/l =0.02$,
$d_\alpha/l =0.1$, $\alpha_{\rm min}=10^{-3}$~s$^{-1}$,
$\alpha_{\rm
max}/\alpha_{\rm min}=2.2\; 10^4$, $\alpha_2/\alpha_{\rm
min}=67$
(i.e., $W_2\simeq 4 k_{\rm B}T$), for $\Omega=1.0\;10^5$~s$^{-1}$
and
$\Omega=1.0\;10^2$~s$^{-1}$. In this system, $v_0\simeq 157$~nm/s
and
$f_{\rm s} = 0.33$~pN. (b) Average attachment time to a filament
$\tau$
as a function of
the normalized force $f_{\rm ext}/f_{\rm s}$,
where again $f_{\rm s} = 0.33$~pN
(with the exception of the
case for
$\Omega=0$ s$^{-1}$ for which the stall force vanishes). Same
model as
in Fig.\ref{f:resB}(a) above, but
$\alpha_2/\alpha_{min}=1.63\;10^5$
(i.e., $W_2\simeq 12k_{\rm
B}T$).}
\label{f:resB}
\end{figure}

Figure \ref{f:resA} shows the average
attachment time $\tau$ as a
function of the externally applied force for
the ISS model for two
different values of the excitation rate $\Omega$
which is an
increasing function of the ATP concentration.  The maximal
attachment
time is independent of $\Omega$ and occurs in the absence of
applied
forces.  In this regime, the system is processive and spends
an
important amount of time attached.  The time $\tau$ decreases if
an
external force which opposes motion is applied.  At stalling
conditions,
the time of attachment has dropped by almost two orders of
magnitude: it
decreases by several orders of magnitude for larger
forces.  The situation
for the DS model is different, as shown in
Fig.\ref{f:resB}.  Forces up to
10-20 times larger than the stall
force are necessary for reducing the
bound time by a factor of $10^2$
for $\Omega=10^2$ s$^{-1}$, for
$\Omega=10^5$ s$^{-1}$ the reduction
does not exceed a factor of $2$ even
for large forces.  The reduced
sensitivity of the processivity to external
forces in the DS model is
due to the fact that the $x$-independent
detachment rate $\alpha_2$
governs this regime.
Our calculations show that
the processivity in
the absence of external forces decreases for increasing
$\Omega$
(cf.~Fig.\ref{f:resB}(a)).  This effect occurs in both models but
is very
significant in the DS model.  For the case of large $\Omega$ we
find
in the DS model a weak dependence of the attachment time as a
function
of tangential load.  Figure \ref{f:resB}(b) exhibits the behavior
of the
DS model for the same parameters as in Fig.\ref{f:resB}(a), but
with
$\alpha_2 \ge \alpha_{\rm max}$ (i.e., $W_2 \ge W_1$).  The
same
general trends are visible, but the excitation-rate dependence
is
strongly increased.  The processivity and force sensitivity are
totally
lost for high excitation rate
$\Omega$.


\section{Discussion and
Conclusion}

Our study shows that the processivity of a molecular motor
is
influenced by an externally applied force tangential to the
filament.
This mechanism applies to all situations in which a molecule
moves along a
linear structure. The details of the initial condition
$P^{(0)}_i$ will not
be relevant if the motor undergoes on average
several cycles and moves a
few periods before detaching, like in the
case of kinesins. However, for
non processive motors, knowledge of the
probability $P^{(0)}_i$ is required
for making detailed predictions.

The ISS model suggests that for
double-headed motors the time of
attachment is particularly sensitive to
the presence of tangential
forces. Indeed, in the presence of
tangential
forces, the motor is more likely to be found
in high-energy
configurations, where the detachment
rate is large. The DS model
demonstrates a
different behavior. In this case, the system spends an
important
fraction of time in
the second state with a
conformation-independent detachment rate
$\alpha_2$ that dominates the
detachment probability and sets the
order of magnitude of the lifetime
$\tau$. Therefore, the time of
attachment depends only weakly on the
applied force.

It has been observed experimentally that the time of
attachment of
kinesin molecules to microtubules decreases as a load is
applied
tangentially to the filament~\cite{copp97}.  These observation
are
consistent with our calculations corresponding to the ISS
model
(Fig.\ref{f:resA}).  Equivalently, calculations of the mean
run
length based
on the ISS model are fully consistent with the results
of
Ref. \cite{schn00}, (Fig.\ref{f:resA}, inset). As a result,
kinesins
lose their
processivity before their direction of motion is
reversed at stalling
conditions and in general reverse motion cannot be
detected.
Furthermore, the attachment time of double-headed kinesins
increases by
about a factor of eight when ATP concentration is decreased at
constant load of $1$pN from
$2$mMol to $5\mu$Mol \cite{schn00}. With the
choice of Fig. 1(a)
we find an increase of the attachment time by a factor
of 4.5
when decreasing $\omega$ by a factor that corresponds to the change
in ATP concentration.
If we slightly increase the
width of the detachment region $\alpha_{max}$
this factor can be further
increased to match the observed ratio.
Our calculations predict that
single-headed kinesins,
which move processively in the absence of external
forces
\cite{okad99}, should in contrast to double-headed kinesins
remain
processive in the presence of tangential
forces. Therefore, the reversal of
the direction of motion could be
observable under loading conditions. It is
further gratifying to
remark that the DS model is successful in explaining
the different
processivities of KIF1A (Fig.\ref{f:resB}(a)), single-headed
conventional
kinesin and myosin II (Fig.\ref{f:resB}(b)): the $W_2$ binding
energy of
KIF1A
has to differ by $8 k_BT$ from that of myosin and by $4
k_BT$ from
that of kinesin. The $4 k_BT$ corresponds indeed to the
difference
measured in \cite{okad00}. In the myosin~II case, processivity
is
totally lost and the calculated behavior in the absence of ATP is
fully
consistent with known experimental data, with and without
external force
\cite{mars82,ishi93,nish95}.

\acknowledgments

We thank D. Nelson both for
communicating to us his work with D.
Lubensky on a different but related
problem, and for illuminating
discussions. AP also thanks D. Nelson for the
hospitality at the
Condensed Matter Theory Group, Harvard University.  We
are grateful to
S. Camalet and K. Kruse for useful discussions. AP's work
was
partially supported by the Deutsche Forschungsgemeinschaft contract
No.
850/4-1. LP's work has been performed within a joint cooperation
agreement
between Japan Science and Technology Corporation (JST) and
Universit\`a di
Napoli \lq\lq Federico II\rq\rq, and with the partial
support of a Chaire
Joliot de
l'ESPCI.

\end{document}